\documentclass[12pt,preprint]{aastex}

\usepackage{natbib}
\usepackage{graphicx}
\bibpunct{(}{)}{;}{a}{}{,}

\newcommand{\lina}{\ion{Fe}{1}~$\lambda$6302.5~\AA}

\newcommand{\linc}{\ion{Fe}{1}~$\lambda$15648~\AA}

\newcommand{\paperiii}{SA05}
\newcommand{\los}{line-of-sight}
\newcommand{\itm}[1]{item~\#~\ref{#1}}
\shorttitle{A topology for the penumbral magnetic fields}
\shortauthors{S\'anchez Almeida}

\begin{document}
   \title{A topology for the penumbral magnetic fields}

    \author{J.~S\'anchez~Almeida}
    \affil{Instituto de Astrof\'\i sica de Canarias, 
              E-38205 La Laguna, Tenerife, Spain}
   \email{jos@iac.es}

\begin{abstract}
	We describe a scenario for the sunspot magnetic
field topology that may account for
recent observations of upflows and downflows in 
penumbrae. According to our conjecture,
short narrow magnetic loops fill the penumbral volume. 
Flows along these field lines are responsible
for both the Evershed effect and the convective
transport.
This scenario seems to be 
qualitatively consistent with most existing observations,
including the dark cores in penumbral filaments
reported by Scharmer et al. Each  bright filament with
dark core would be a system of two paired convective rolls with the
dark core tracing the lane where the plasma  sinks down.
The magnetic loops would have a hot footpoint
in one of the bright filament and a cold 
footpoint in the dark core.
The scenario also fits in most of our theoretical
prejudices (siphon flows along field lines, presence
of overturning convection, drag of field lines by downdrafts,
etc). 
If the conjecture turns out to be  correct, 
the mild upward and
downward velocities observed in penumbrae must
increase upon 
improvement of the current spatial
resolution. This and other observational
tests to support or
disprove the proposed scenario are put forward.
\end{abstract}
\date{December 21, 2004}
\slugcomment{Original version submitted to ApJ on December 21, 2004, but never published}
%

\section{Introduction\label{introduction}}

The spectral line asymmetries observed in sunspots may arise
from small scale variations of the 
sunspot magnetic field. When the size of 
the inhomogeneities is small enough several of them
are integrated along
any \los ,
thus generating asymmetries \citep{san96}. In this
picture 
the large scale variation of the sunspot magnetic 
field determines the symmetric part of the
line shapes whereas
the optically-thin micro-structure
yields asymmetries.

Using this MISMA\footnote{
The acronym MISMA stands for 
MIcro-Structured Magnetic Atmosphere, and it
was coined by \citet{san96} to describe 
magnetic atmospheres having optically-thin
substructure.} 
framework,
\citet[][hereinafter \paperiii]{san04b} 
carries out a systematic fit of 
all Stokes profiles\footnote{We use   
Stokes parameters to characterize
the polarization of a spectral line; $I$ for the intensity,
$Q$ and $V$ for the two independent types
of linear polarization, and $V$ for the
circular polarization. The 
so-called Stokes profiles are the set
of these Stokes parameters versus 
wavelength for a particular spectral line.}
of \lina\ and \lina\ 
observed with fair angular resolution ($\sim$1\arcsec ) 
in a region with a medium size sunspot 
(radius~17~Mm).
Some 10130 independent
fits allow to characterize the sunspot
magnetic field.
The resulting semi-empirical model sunspot 
provides both
the large scale magnetic structure as well as the
small scale properties of the micro-structure.
As far as the large scale properties is concerned, 
the sunspot magnetic field agrees
with previous measurements
\citep[e.g.,][]{bec69c,lit90,lit93}. 
A roughly
vertical umbral magnetic field becomes progressively
more inclined toward the sunspot border, 
where it approaches the horizontal.
Mass flows 
are absent in the umbra. They
appear in the penumbra and 
increase toward the penumbral border.
On top of the regular expected large scale behavior,
the inferred small scale structure of the magnetic 
fields and flows is new and unexpected. 
Some 40\% of the
magnetic field lines bend over and return to the 
photosphere within the penumbra. This return of magnetic
flux and mass toward the solar interior occurs 
throughout the penumbra, as opposed to previous 
claims of bending over and return  at the penumbral 
border or beyond \citep{rim95,wes97,tri04}.

Counter-intuitive as it may be, having 
field lines pointing up and down all over the
penumbra does not seem to conflict with 
existing observations. They have not been 
inferred before because the upflows and downflows and the
positive and
negative polarities coexist in scales 
smaller than the typical resolution elements 
(1\arcsec ). Most measurements assume 
resolution element with uniform properties, and the
mean values thus derived tend to show upward pointing
magnetic field lines.  These measurements often
present serious problems of self consistency 
(e.g., non-parallelism between magnetic field lines and
flows, \citealt{are90};
violation of the conservation of magnetic flux, \citealt{san98a};
non-parallelism between continuum filaments and 
magnetic field lines, \citealt{kal91}).
On the other hand, the existence of field lines
that bend over and return 
does not clash with any theoretical prejudice.
Rather, such pattern of motions is expected
from first principles \citep[e.g.,][and \S~\ref{observations}]{hur00}.

This paper completes and complements \paperiii . 
The existence of an ubiquitous return of magnetic
flux, together with many
results from the literature, are 
assembled to offer a
plausible scenario for the penumbral magnetic field 
topology. Such exercise  to piece together and synthesize
information from different sources is confessedly 
speculative.
We will not provide a self-consistent MHD model for 
the penumbral structure. However, the exercise is needed.
First, it  explains how the results in \paperiii\ 
are qualitatively consistent with most 
existing observations. 
Second, the scenario
may inspire MHD numerical modeling 
with reasonable chances of
reproducing the existing observations.
Finally, despite the amount of papers on
penumbrae, our understanding of the penumbral 
phenomenon is far from satisfactory. 
We offer a solution with an alternative twist.
Even if the proposal  turns out to be incorrect, it
would stir a debate with a few
novel ingredients which may eventually be important 
to understand the true penumbral structure.

We begin by 
summarizing several results on the penumbral 
structure used to set up the stage
(\S~\ref{observations}). The scenario is put forward in
\S~\ref{scenario},
and its predictions are qualitatively 
compared with observations in \S~\ref{qua_obs}.
Similarities and differences between
this scenario and existing models for the
penumbral structure and the Evershed flow are
analyzed in \S~\ref{theory}. Finally, we 
suggest several observational tests to confirm
or disprove our conjecture (\S~\ref{conclusions}).

\section{Commented references on the penumbral structure}
\label{observations}

	The bibliography on the magnetic structure
of penumbrae is too extensive to quote\footnote{The NASA Astrophysics
Data System provides about one thousand papers
under the keyword {\em penumbra}. Sixty of them
were published during the last year.}.
We refrain from giving an overview of the
field and refer to recent reviews:
\citet{sch91,tho92,sol03,bel04,tho04}. Only a few selected references
are introduced and discussed to provide the framework of our work.
They are either complementary to the \paperiii\ findings or
of direct relevance to our interpretation. 
Most of them are observational results but some involve
theoretical arguments.
The selection is obviously biased in the sense 
that some observations often bypassed are emphasized
here, and vice versa. However, 
to the best of our knowledge,
no potentially important result has been excluded.

\begin{enumerate}
\item \label{best}
The best penumbral images have an angular resolution
of the order 0\farcs 12 (or 90~km on the Sun).
They show many features at the
resolution limit implying the existence
of unresolved structure. For example,
the power spectrum of the penumbral images have
power up to the instrumental cutoff \citep{rou04},
and the width of the narrower penumbral filaments is
set by the resolution of the observation 
(\citealt{sch02}; see also Fig.~\ref{narrow}).
This interpretation of the current observations
should not be misunderstood. The penumbrae have
structures of all sizes starting with the penumbra
as a whole. However, the  observations show that
much of its observed structure is at the resolution
set by the present technical limitations and,
therefore, it is expected to be unresolved.
\begin{figure}
\plotone{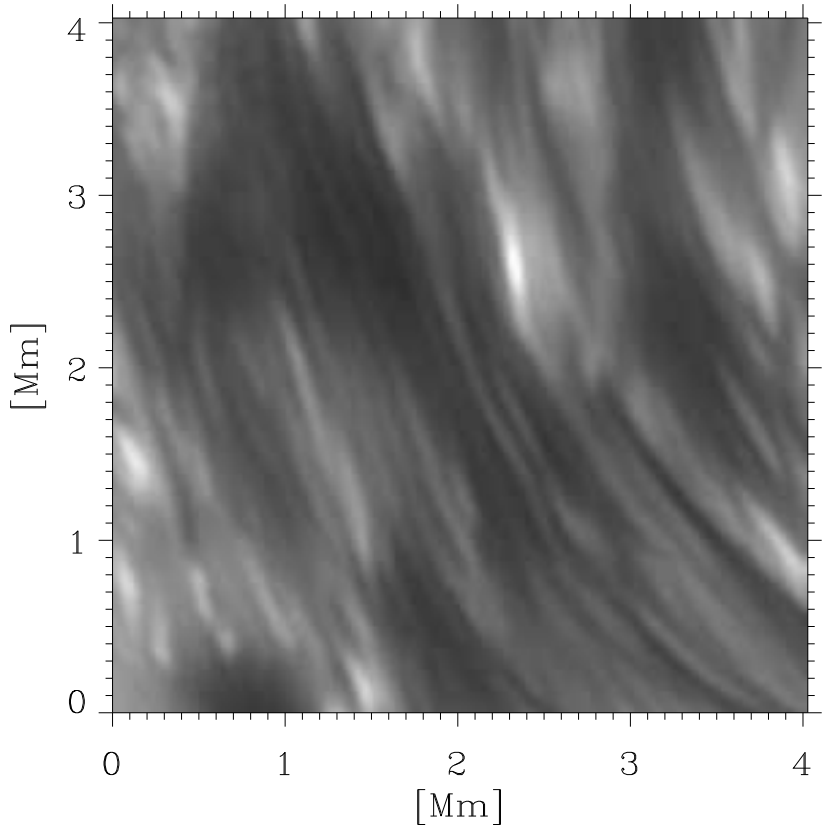}
\caption{Small fraction of the penumbra 
	observed by \citet{sch02}. Note the
	narrowness of the filaments, and their
	large aspect ratio (length over width).
	The spatial scales are in Mm, and the 
	angular resolution of the image is of the order
	of 0.09~Mm.}
\label{narrow}
\end{figure}

\item \label{cores}
The best penumbral images show 
{\em dark cores in sunspot penumbral filaments}
\citep{sch02}. 
We prefer to describe them as  dark filaments
surrounded by bright plasma.
This description also provides a fair account
of the actual observation (see Fig.~\ref{coresfig}), 
but it 
emphasizes the role of the dark core. Actually,
dark cores without a bright side are 
common, and the cores seldom emanate from
a bright point (Fig.~\ref{coresfig}).
\begin{figure}
\plotone{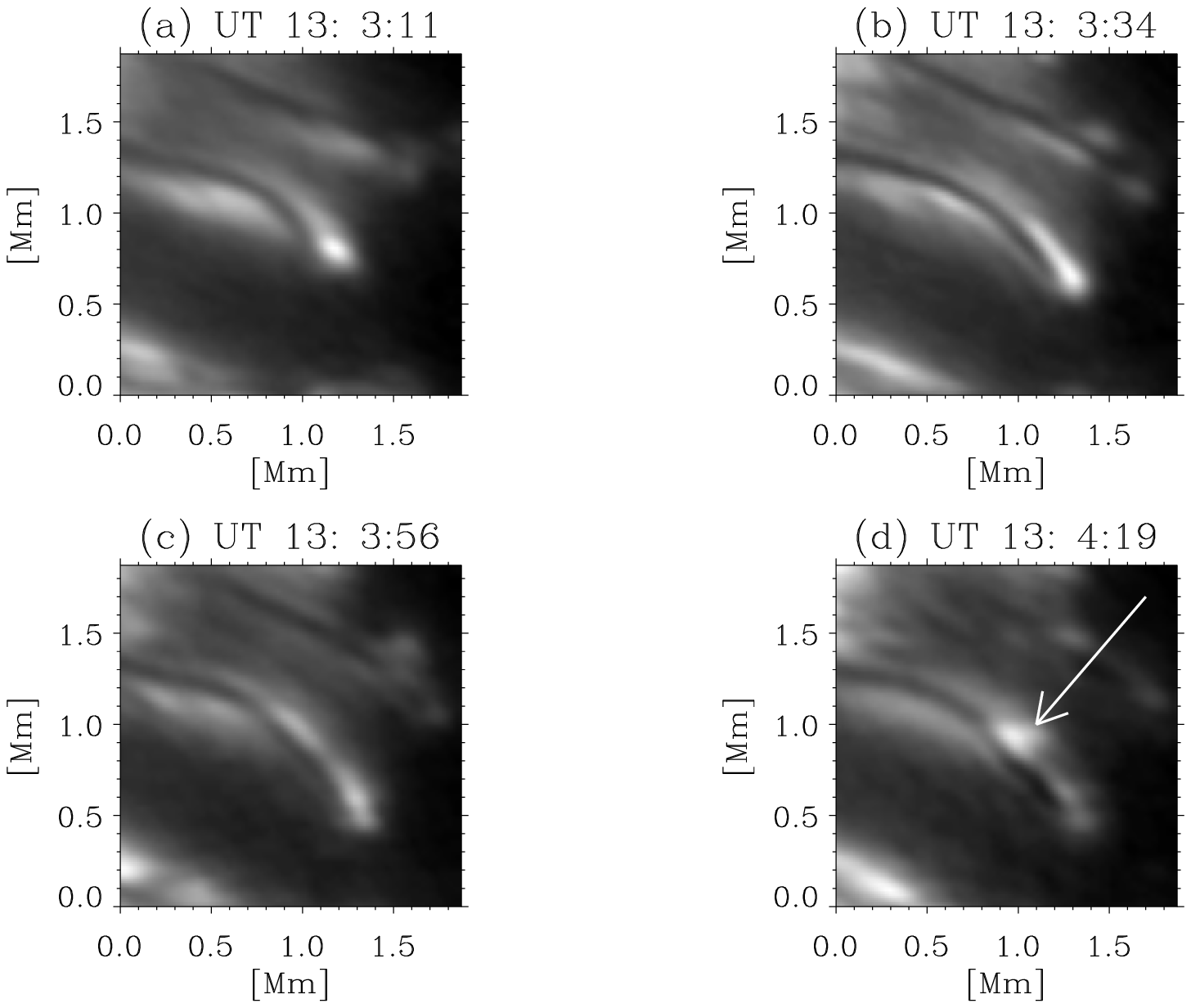}
\caption{Time evolution
of one of the {\em dark cores in penumbral filaments}
discovered by \citet{sch02}. (The
UT of observation is marked on top of each snapshot.)
Note that one of the
bright sides is partly missing  in (c) and (d). Note also
that the bright points are not on the dark filament
but in a side. These two
properties are common. The arrow indicates the
emergence of a new bright point in a side of
the pre-existing dark filament.}
\label{coresfig}
\end{figure}
The widths of the dark core and its bright 
boundaries
remain  unresolved,
although the set formed by a dark core 
sandwiched between two bright filaments
spans some 150-180~km across.

%

\item \label{correlation}
	There is a local correlation 
	between penumbral brightness
	and Doppler shift, so that bright features
	are blueshifted with respect to the dark
	features \citep{bec69c,san93b,joh93,sch00b}. The same
	correlation exist both in the limb-side
	penumbra and the center side penumbra, 
	a fact invoked by \citet{bec69c} to conclude
	that vertical motions rather than horizontal
	were responsible for the correlation. 
	A positive correlation between vertical velocity
	and intensity is characteristic of the 
	non-magnetic granulation.
	The fact that the same correlation
	is also present in penumbrae is suggestive
	of a common origin for the two phenomena, namely,
	convection.
\item\label{this}
    At a constant distance from the sunspot center, the 
    limb-side penumbra and center-side penumbra are darker than the
    rest of the penumbra. This observational effect indicates 
    that the  bright features are more opaque than the dark
    ones \citep{sch04}.
\item\label{light_bridge}
	Light bridges (LB) are often formed by a dark filament
	surrounded by chains of BPs. See, e.g., Fig. 6
	in \citet{lit04}. The arrangement resembles the
	dark cores in penumbral filaments (compare
	the two features in Fig.~\ref{light}). 
	The (low spatial
	resolution) magnetic field vector in LBs is also similar
	to that in penumbrae.
	It tends to be weak, horizontal,
	and aligned along the bridge \citep{lek97}.
	LBs present a positive correlation between upflows and
	brightness \citep{rim97}, as penumbrae do (\itm{correlation}).
	Moreover, the bright points surrounding the dark channel
	seems to be elevated \citep{lit04}, as the bright
	penumbral filaments (\itm{this}).
\begin{figure}
\plotone{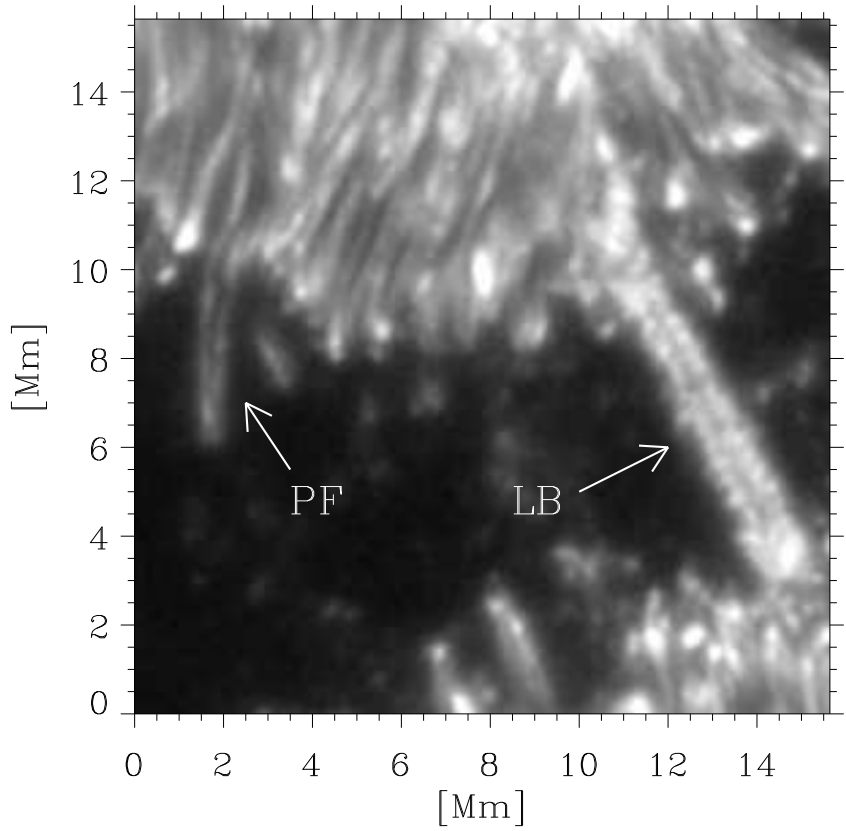}
\caption{Light bridge (LB) showing a dark channel
	surrounded by two bright filaments. It looks like a scaled
	up version of the dark cores found in penumbrae 
	(see the one marked with PF). Image obtained with the
	Swedish Vacuum Solar Tower \citep[SVST; ][]{sch85} 
	equipped with adaptive
	optics and frame selection. 
	}
\label{light}
\end{figure}
\item There is a local correlation between the 
	magnetic field inclination and horizontal velocity.
	The largest velocities are associated with
	the more horizontal fields
	\citep[e.g.][]{tit93,sta97}.

\item \label{horizontal}
	The large horizontal motions occur in the 
	dark penumbral filaments \citep[e.g.][]{rue99,pen03}.

\item \label{contradict}
	The observations on the correlation
between magnetic field strength and brightness
are contradictory. Some authors find the 
strongest field strengths associated with the
darkest regions and vice versa (c.f., \citealt{bec69c} 
and \citealt{hof94}).
However, the condition of mechanical
balance between a hot component and a cold component favors
a positive correlation, i.e.,  strongest
fields occurring in the darkest (coldest) regions. 
One expects a deficit of gas pressure in low temperature
regions which should be balanced by the increase of magnetic
pressure, and so, the increase of magnetic field
strength. The magnetostatic equilibrium along field
lines also favors dark regions with large field strengths
as argued by, e.g., \citet{san01c}.

\item  \label{roll1}
	The stable sunspots are surrounded by a large
	annular convection cell called moat \citep{she72}.
	The moat presents an outflow which, contrarily to the 
	commonly held opinion,
	has  both radial and tangential velocities  
	(Title 2003, private communication;
	\citealt{bon04}). The tangential component
	is well organized so that it sweeps the plasma 
	toward radial channels, creating a
	velocity pattern that resembles the convective
	rolls by \citet{dan61}. Compare Fig.~7
	in \citet{bon04} or Fig.~11 in
	\citet{bov03} with the rolls in Fig.~9a of  \citet{gar87}.

\item	\label{roll2}
	Actually, theoretical arguments indicate that
	the convective roll pattern seems
	to be the preferred mode of magneto-convection 
	for nearly horizontal magnetic fields 
	\citep{dan61,hur00}. The rolls have their axes
	along the magnetic field lines.
\item \label{drag}
	The magnetic field lines can be dragged by the 
	downdrafts of the granulation. 
	Termed as {\em flux pumping} mechanism,
	this drag is modeled and studied by \citet{wei04}
	to show that the vigorous sinking plumes of the
	granulation and mesogranulation easily pumps down 
	magnetic fluxtubes outside the penumbra.
	It is conceivable that 
	the same pumping by sinking 
	plasma also operates
	in the magnetized penumbra. It would tend to sink
	down cool plasma.

\item \label{ncp}
	The polarization  of the spectral lines emerging from 
	any sunspot has asymmetries, i.e., they have to be
	produced in atmospheres with several velocities 
	and magnetic fields per resolution element 
	\citep{bum60,gri72,gol74,san92b}. The spatial resolution
	of the spectro-polarimetric observations is seldom
	better than 1\arcsec or 725~km. However,
	the unresolved structure producing asymmetries
	has to be  smaller than this size.
	The spectral lines create 
	net circular polarization (NCP), i.e., the asymmetry
	of the Stokes~$V$ profiles is such that the
	integrated circular polarization of any spectral
	line is not zero. NCP can only be produced by gradients
	along the \los\ and, therefore, within a range
	of heights smaller that the region where
	the lines are formed (say, 150~km).
	The NCP follows several general rules found by
	\citet{ill74a,ill74b} and \citet{mak86}, and
	summarized in \citet[\S~4]{san92b};
	it has the sign of the sunspot polarity
	and the maximum occurs at the limb-side penumbra where
	the Stokes $V$ profiles are extremely asymmetric
	with three or more lobes. Recently, \citet{sch02b}
	have found that the highly magnetic sensitive
	\linc\ line does not follow the general rules.

\item \label{cooling}
\citet{sch99} estimate the length of a bright
filament produced by a hot flow along a 
magnetic fluxtube. It cools down as it radiates away and so, 
eventually, the plasma in the fluxtube becomes dark
and transparent.
An isolated loop  would have 
a bright head whose length $l$ is approximately set  by 
the cooling time of the emerging plasma $t_c$ times the velocity
of the mass flow along the field lines $U$ (e.g., \citealt{sch99}),
\begin{equation}
l\sim U t_c.
\end{equation}
The cooling time depends on the diameter of the tube $d$,
so that the thiner the tube the  faster the cooling \citep[e.g.,][]{sti91}.
\citet{sch99} work out the cooling time for a hot
fluxtube in a penumbral environment, and it turns 
out to be of the order of 
\begin{equation}
t_c\sim 30~{\rm s}~(d/200~{\rm km})^{1.5}.
\end{equation}
For reasonable values of the Evershed flow speed of the
emergent plasma 
(U$\sim$ 5~km~s$^{-1}$),
\begin{equation}
l\sim 150~{\rm km}~(d/200~{\rm km})^{1.5}.
\end{equation}
Then the aspect ratio
of the hot footpoint (i.e., the ratio between the length
and the width) is 
weakly dependent on the diameter, i.e.,  
\begin{equation}
l/d\sim 0.8~ (d/200~{\rm km})^{0.5}.
\end{equation}
Such aspect ratio does not represent a filament. Filaments 
must have $l/d >> 1$,  
consequently, the cooling of hot plasma
moving along field lines does not give rise
to the kind of narrow observed  filaments
(see Fig.~\ref{narrow}).
We are forced to conclude that the penumbral 
filaments cannot be tracing individual fluxtubes.
If arrays of fluxtubes form the filaments, they
must be arranged with their
cold and hot footpoints aligned to
form the observed structures.

\item \label{up_down} 
The sunspots seem to have upward motions in the inner penumbra and
downward motions in the outer penumbra
\citep[e.g.][]{rim95,sch00,bel03b,tri04}.
However,
this velocity pattern is inferred 
assuming uniform velocities in the resolution elements,
an hypothesis inconsistent with other existing 
observations
\citep[e.g.][]{bum60,wie95}.

\end{enumerate}

\section{Scenario for the 
	small-scale structure 
	of the penumbra}\label{scenario}

\begin{figure}
\plotone{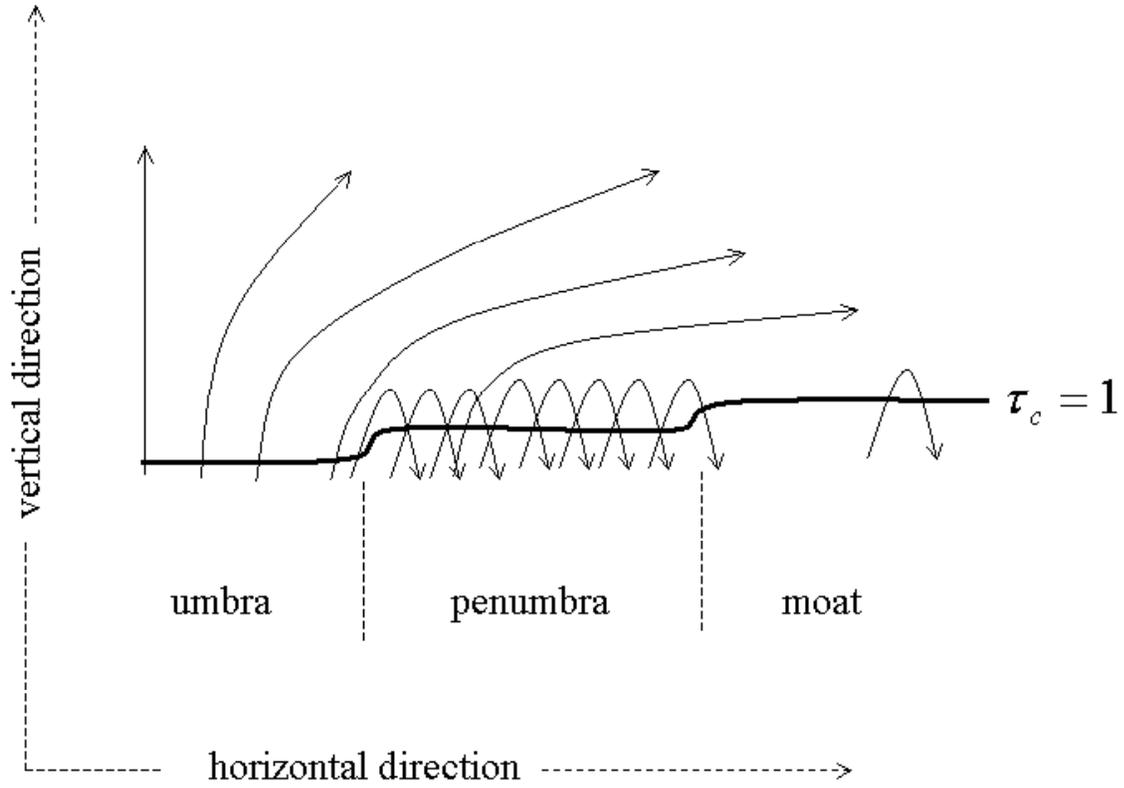}
\caption{Cartoon sketching the scenario for the penumbral
	magnetic field topology put forward in the paper.
	Magnetic field lines, represented as solid
	lines with arrow heads, return to the sub-photosphere
	in the entire penumbra. The symbol $\tau_c$ stands
	for the continuum opacity so that the thick solid line
	marked as $\tau_c=1$ represents the base of the photosphere.
	}
\label{cartoon2}
\end{figure}

\begin{figure}
\plotone{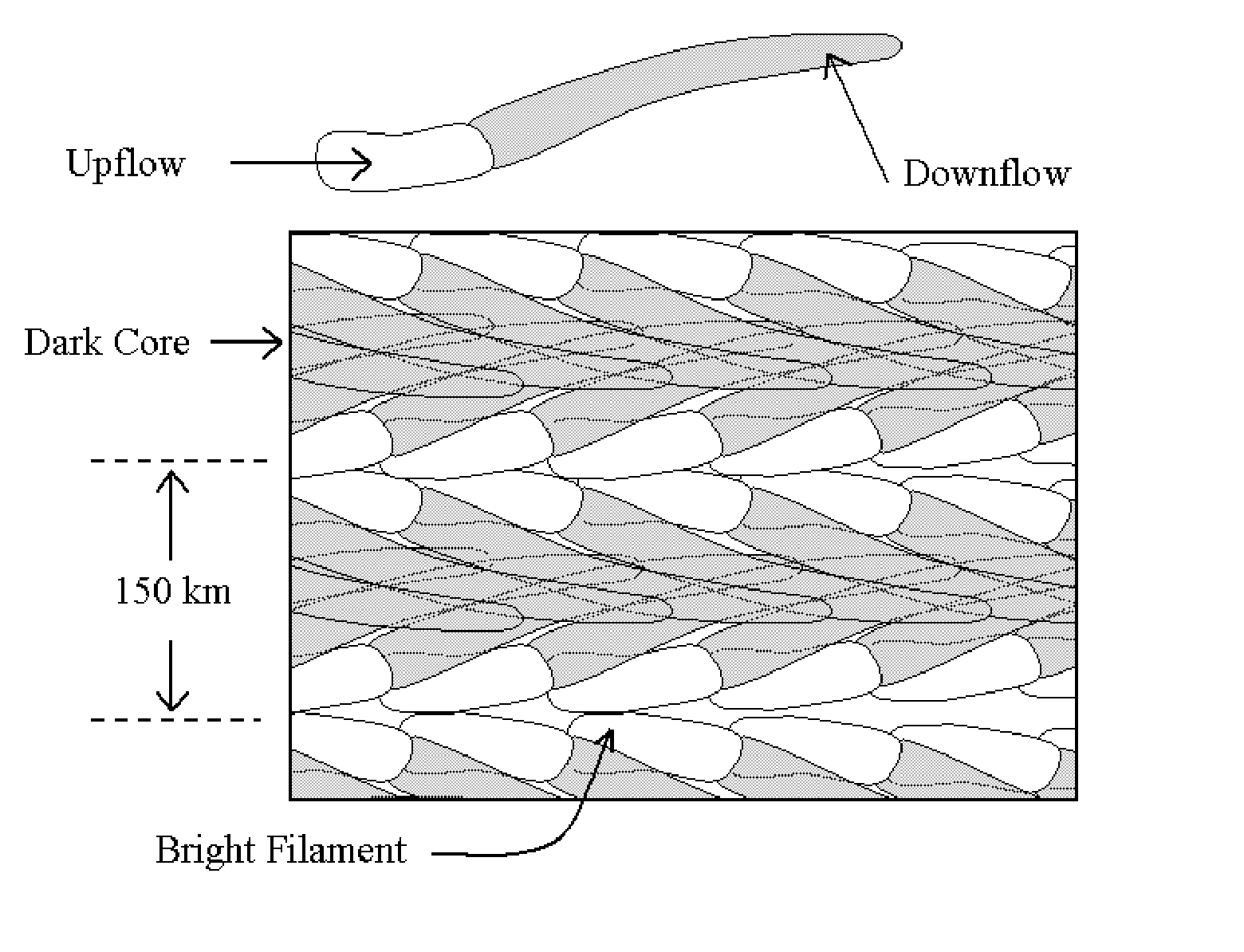}
\caption{A view from the top of a small
	portion of penumbra (see the scale on the left hand side 
	of the cartoon).  Small loops like the one on top
	of the figure are averaged in our resolution element 
	(the rectangle).
	They are so thin that various loops overlap
	along the \los .
	The loops are arranged with the downflowing footpoints
	aligned forming a dark core. The hot upflows
	feeding a dark core give rise to two bright filaments.
	The Evershed flow is directed along the field lines
	toward the right.}
\label{cartoon1}
\end{figure}

The penumbra may be made out of short narrow shallow  magnetic 
loops which often return to the sub-photosphere 
within the sunspot boundary (Fig.~\ref{cartoon2}). 
One of the footpoints is hotter than the other
(Fig.~\ref{cartoon1}).
The matter emerges in the hot foot point, radiates away, cools down,
and returns through the cold foot point.
The ascending plasma is hot,  dense, and  slowly moving.
The descending plasma is cold, tenuous, and  fast moving.
The motions along magnetic field lines
are driven by magnetic field strength
differences between the two footpoints, as
required by the siphon flow mechanism.

\begin{figure}
\plotone{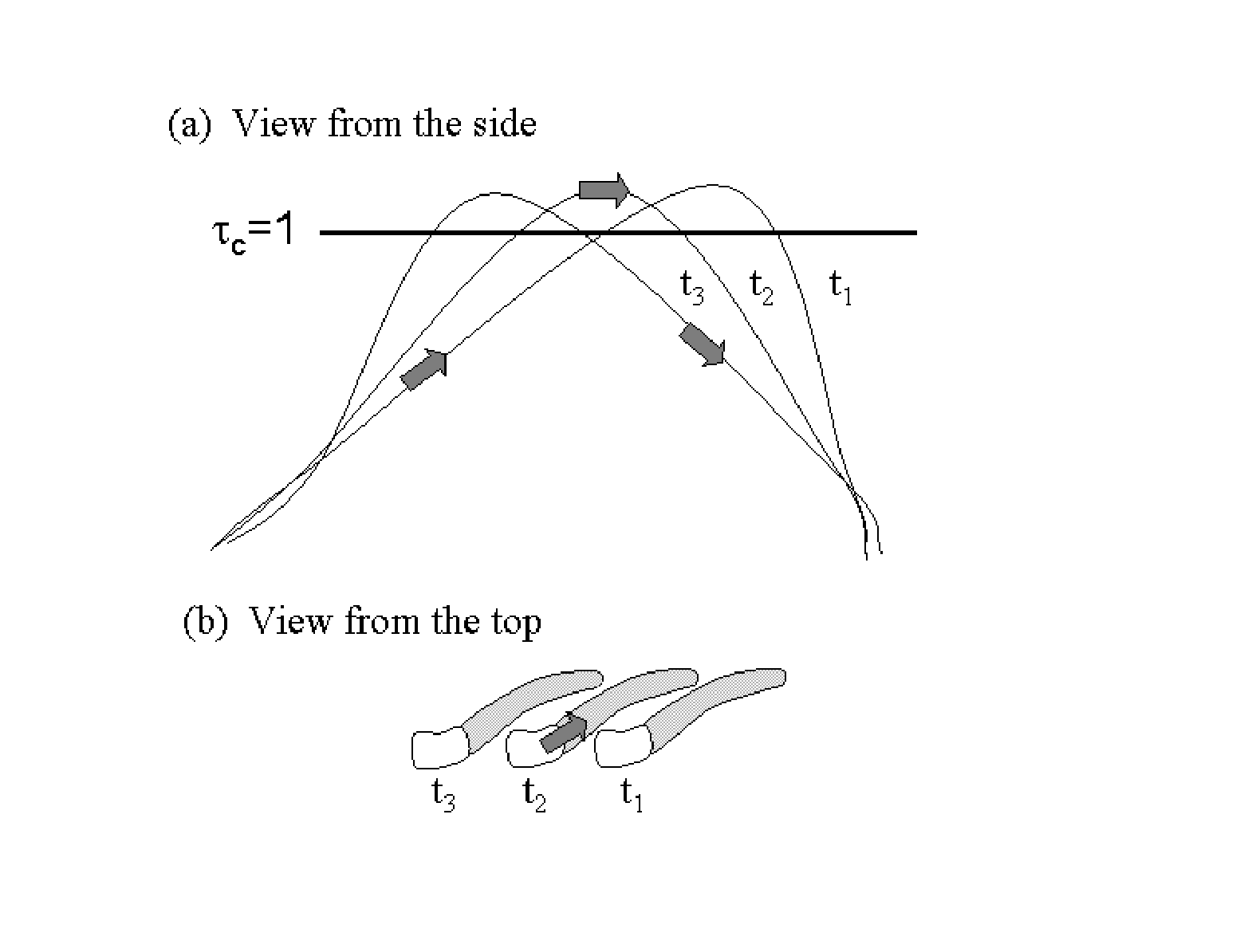}
\caption{Cartoon showing the time evolution 
	of a single field line with time $t_1 < t_2 < t_3$
	(the thin solid lines). 
	(a) View from the side, with the
	vertical direction pointing upward. The thick solid line
	corresponds to base of the photosphere (continuum opacity
	equals one). The Evershed flow is directed
	to the right.
	The thick arrows represent a parcel of fluid with the 
	tip pointing in the direction of the plasma motion.
	(b) View from the top of the same three instants.
	(Compare with Fig.~\ref{cartoon1},
	showing many field lines at a given time.) It only shows
	the fluid parcel at $t_2$ since it remains below the 
	photosphere at $t_1$ and $t_3$.
	}
\label{cartoon3}
\end{figure}

In addition to the large velocities along field lines,
the cold footpoint sinks down while
the hot footpoint rises.
This produces a backward and sideways displacement of the
visible part of the loop (see Fig.~\ref{cartoon3}).  
The sink of the cold leg could be induced by the drag 
of downflowing plumes or downflowing sheets  existing in the
observed dark cores (Fig.~\ref{coresfig}).  
The non-magnetic downflowing plumes can 
pump down magnetic fields lines
as modeled by \citet{wei04}. We are proposing
a magnetically modified version of this mechanism.
As it happens with the non-magnetic convection,
upflows are indirectly driven
through mass conservation by displacing
warm material around the downdrafts
\citep{stei98,ras03}. The uprising hot material tends
to emerge next to the downflows. 
The properties of the loops (length, width, speed,
and so on) should
change depending on the 
position on the penumbra, but they would
tend to be narrow and so optically-thin
across field lines.
Since they are optically-thin,
one does not detect individual loops but assembles
of them interleaved along the \los.
The cold legs of many different loops should be
identified with the dark cores found by 
\citet{sch02};
compare Figs.~\ref{coresfig} and \ref{cartoon1}. 
The upflows of many loops account for be bright
penumbral filaments.
In this
scenario, magnetic field lines are not exactly aligned
with the penumbral bright and dark filaments but the field lines
diverge from the bright filaments and converge
toward the dark filaments.
The mean field is radial, though.
This arrangement occurs in a scale not yet resolved
by the present spectro-polarimetric observations.

According to our scenario,
bright filaments are associated with
fields having the polarity of the sunspot,
mild upflows,
low field strengths,
and they are
opaque\footnote{They are both dense and hot,
two factors that augment the opacity
of a structure.}. On the contrary,
dark filaments are associated with 
fields whose polarity is opposite to the
sunspot, have intense flows, high field strength, 
and they are more transparent than the bright filaments.
\section{Qualitative comparison with observations}\label{qua_obs}
The scenario 
described in \S~\ref{scenario} seems to
fit in the  observations 
in \S~\ref{observations} plus the
return of magnetic flux and mass flow in \paperiii .
This section points out why.
The items mentioned along the text refer
to those numbered in \S~\ref{observations}.

 The model penumbra is covered by narrow loops whose length is typically 
smaller than the penumbral size (Fig.~\ref{cartoon2})
and whose widths are not spatially resolved, in agreement
with \itm{best}. Consequently, magnetic field lines
emerge and return to the sub-photosphere over the entire penumbra, 
as inferred in \paperiii . The loops have a
hot footpoint with upward motion and a cold footpoint with 
downward motion, in agreement with the local correlation between
brightness and upward velocity observed 
in penumbrae
(\itm{correlation}). The downflows are expected to be faster 
than the upflows
since they are accelerated by the magnetic field  difference
between the two footpoints, an image that also fits in
well  the observations showing the largest  
velocities to be associated with the dark penumbral
components (\itm{horizontal}) and with 
the strongest magnetic field strengths (\paperiii ; \itm{contradict}).

We identify the  dark cores found
by \citet[][\itm{cores}]{sch02}
with cold footpoints of many loops, as 
sketched in Fig.~\ref{cartoon1}.
They  trace downdrafts
engulfing the cold footpoints
of the loops (\itm{drag}).
The bright filaments
around the dark cores would be naturally explained 
by the presence of the downflows, as it happens 
with the enhanced brightness at the borders
of the granules in non-magnetic convection.
When the downdrafts reach 
sub-photospheric depths, they displace hot material
next to them. This hot material tends to rise to satisfy
mass conservation, and it produces an upflow
of hot material around the downdrafts \citep{ras95,ras03,stei98}.
The same mechanism
would produce the uprise of hot (magnetized) material
around the cores, forming two
bright filaments outlining each dark core
(\itm{cores}; Fig.~\ref{coresfig}). The hot
magnetized material would eventually cool down and
sink into the
downdraft to re-start the process.
In other words, a
dark core would be the downdraft of two paired
convective rolls, 
resembling the convective
rolls proposed by Danielson  
(\itm{roll1}). In this
case, however, the plasma also has a large
velocity component
along the field lines. Note that this hypothetical
convective rolls reproduce the expected mode of 
convective transport 
in highly inclined magnetic fields (\itm{roll2}).
Moreover, a pattern of motions similar to these
convective rolls occurs in the moat surrounding
the sunspot  (\itm{roll1}), and it is conceivable that
it continues within the sunspot.
Light bridges like that  in Fig.~\ref{light} would
be a scaled up version of the doubly convective
rolls that we are proposing as building blocks
for the penumbral structure (\itm{light_bridge}). They 
have a dark lane
surrounded by two chains of bright points forming filaments. 
This distribution matches the cartoon in Fig.~\ref{cartoon1}.

The magnetic field lines of
the loops converge toward the dark lanes (Fig.~\ref{cartoon1}),
which would demand an increase of the magnetic
field strength in the cores. This is indeed 
consistent with our scenario, and it  would
be responsible for the gas pressure difference
driving the Evershed flow along the field lines
(see above).

The existence of small scale convective 
upflows and downflows in penumbra does
not contradict the systematic
upward motions in the inner penumbra and
downward motions in the outer penumbra
found by various authors (see \itm{up_down}).
Most techniques employed so far 
assume uniform velocities in the resolution element.
When spatially unresolved 
upflows and downflows 
are interpreted as a single
resolved component, the {\em measured}
velocity corresponds to an ill-defined mean 
of the actual velocities. 
The contribution of upflows and downflows
to the mean is not proportional to the mass going up and down.
It depends on the physical
properties of the upflows and downflows, as well
as on the method employed to measure. The
mean vertical flux of mass  inferred by \paperiii\
is null. However, the average is biased so that
if shows net upflows in the inner penumbra
and net downflows in the outer penumbra, in agreement
with existing observations\footnote{The effect is similar
to the convective blueshift of the spectral lines formed
in the granulation, which does not imply a net
uplifting of the quiet photosphere.}.

This scenario with overlaying loops with various
velocities and inclinations
satisfies the rules of the net
circular polarization 
described in \itm{ncp}, as
worked out by  \paperiii .

The bright filaments are more opaque than the
dark cores, and they tend to block the cores
when the filaments are observed from a side.
This depression of the dark cores
explains why
penumbrae are slightly darker
in the line along the radial direction from the
solar disk center (\itm{this}).

The length of the bright filaments is not set 
by the cooling time of individual fluxtubes, which
avoids the difficulty posed in \itm{cooling}.
It is given by the length of the dark core, i.e., 
the length of a sheet-like downdraft or 
the length of a radially oriented chain of many 
small downdrafts.

\subsection{Does the model carry enough energy
to account for the penumbral radiative
flux?}\label{not_enough}
The radiative flux emanating from a penumbra $F$ is some
75\% of the solar flux \citep[e.g.][]{sch03f}, which corresponds to
\begin{equation}
F\simeq 5\times 10^{10}~{\rm erg~cm^{-2}~s^{-1}}.
\end{equation}
The need to balance this loss with energy
transported by convective motions 
sets the typical convective velocity $U_z$ 
\citep[e.g.,][]{spr87,stei98}; 
\begin{equation}
F=U_z\rho\alpha \epsilon,
\end{equation}
with $\rho$ the density, $\alpha$ fraction
of atmospheric volume occupied by upward
motions and $\epsilon$ the energy per unit mass
to be radiated away. 
Using typical values,
\begin{equation}
U_z\simeq 1~{\rm km~s^{-1}} 
\bigl[
{\rho\over{2\times 10^{-7}{\rm~g~cm}^{-3}}}
{\epsilon\over{3\times  10^{12}{\rm~erg~g}^{-1}}}
{\alpha\over {0.7}} 
\bigr]^{-1}.
	\label{vert_veloc}
\end{equation}
The density and occupation fraction are representative
of the upflows in the inversion by
\paperiii, whereas $\epsilon$ is taken from
\citet{sch03f}\footnote{
Most of the convective energy is transported
as ionization energy of H ions \citep[see,][]{stei98}.
Consequently, 
$\epsilon$ cannot be much larger that the value 
adopted above.
A maximum is set by the energy corresponding
to one recombination event
per existing H particle,
i.e.,
the H ionization potential
divided by the mass per H particle, which
equals 
some $10^{13}$~erg~g$^{-1}$.
}.
If the transport of energy in penumbrae is due
to convective motions, then there should be 
a sizeable vertical velocity of hot
plasma over the entire penumbra, as indicated
by equation~(\ref{vert_veloc}).
Unfortunately, the vertical velocities for the 
upward moving component in \paperiii\
are one order of magnitude smaller than 
the speed in equation (\ref{vert_veloc}).
(The same happens with the vertical velocities
discussed in \itm{correlation}.)

The above back-of-the-envelop estimate shows that
the upflows and downflows in \paperiii\
do not transport enough mass to account for 
the penumbral radiative 
loss\footnote{
Actually this discrepancy  between required and 
observed velocities  was used
to discard the transport of energy 
by convection in penumbrae 
\citep{spr87}, leading to the concept
of shallow penumbra \citep{sch86b}.}.
A possible way out would be
the existence of a  bias that underestimates
the true velocities. This bias is to be expected
in measurements which 
do not distinguish between unresolved
component, e.g., those leading to the results
discussed in \itm{up_down}
(see Appendix~\ref{appa}).
The reason for a bias in \paperiii\ is more uncertain,
but could be due to a misinterpretation
of bulk upward and and downward motions as
turbulent motions. Actually, 
micro-turbulence in excess of 1~km~ is not
uncommon in \paperiii .

If this bias is present, the observed
vertical velocities in sunspots must increase
upon improvement of the angular resolution,
a conjecture that can be tested
with instrumentation now
available.
Keep in mind, however, that the
true velocity signal
would be severely damped unless the velocity
pattern is spatially resolved 
(Appendix~\ref{appa}). If the scenario
advocated in the paper turns out to be
correct, the detection of sizeable upflows and 
downflows requires a spatial resolution 
comparable with the size of the bright
penumbral filaments with dark cores
found by \citet{sch02}.

\subsection{Why does the low-density plasma
of the cold footpoints sink rather than float?}
We have been arguing by analogy with the
non-magnetic convection, where the
(negative) buoyancy forces in the intergranular
lanes are the driving force. The plasma 
tends to sink down due to its reduced density
as compared to the hot upwelling plasma.
The scenario for the penumbral
convection discussed above does not reproduce
this particular aspect of the granular
convection. The  descending footpoint
has reduced density as compared to
the upflowing footpoint. The density in the
descending leg is lower than in the ascending leg,
and one may think that the descending plasma
is buoyant.
However, 
the density of the cold leg has to be compared to the
local density in the downdraft, which 
can easily be larger than the downdraft
density. Recall that the downdrafts have low
temperature and high magnetic field strengths,
two ingredients that naturally produce
low densities in a
magneto-hydrostatic equilibrium 
\citep[see, e.g.,][]{san01c}.

\subsection{Why are the dark cores radially
oriented?}
According to our conjecture, the dark cores
or penumbral lanes are
not tracing individual field lines. They are faults
in the global structure of the sunspot
where downflow motions are easier.
	 Such discontinuities of the global
magnetic structure would be favored 
if they are aligned with the mean magnetic
field, as it happen with the interchange 
instabilities, which  have an old
tradition in the sunspot literature 
\citep{par75,mey77,mor89,jah94}. 
The dark cores would  
be oriented along the direction of the 
mean penumbral magnetic field.

\section{Qualitative 
comparison with models}\label{theory}

	The scenario that we have assembled 
owes much of its properties to various model
for the penumbral structure and the Evershed
effect existing in the literature.

The model that emerges is very much like the return
flux model of \citet{osh82}, where the field lines
forming umbra and penumbra differ because the former
are open whereas the latter return to the photosphere.
The penumbral field lines are forced to return
to the photosphere due to the expansion of the
umbral fields. 
The return flux is included in our scenario, except that the
return of penumbral field -lines occurs throughout
the penumbra, rather than outside the 
sunspot border. 

The siphon flow model for the Evershed
effect by \citet{mey68} precedes all further modeling  that
assume several magnetic components with different
magnetic fields and velocities  coexisting in
the penumbra. Quoting \citet{mey68},
{\em ... in the penumbra we have
two different sets of magnetic flux tubes side by side.
A small fraction are elements lying very low in the solar
atmosphere and presumably starting with very small
inclination angles. The larger fraction carrying most
of the observed magnetic flux reaches up high into the
atmosphere and would display no flow velocity at all.}
The flows are accelerated by gas pressure
differences between the penumbra and magnetic
concentrations outside the sunspot. In the 
latest version of the model \citep{tho93,mon97},
the loops are fairly high (200--350 km above
$\tau_c=1$) and short (less than 1.5~Mm), and
some returns within the penumbral boundary.
The loops require an external force that keeps the
two foot points anchored to sustain the equilibrium
of the arch. This force has been recently attributed
to the drag of downflowing plumes in the granular
convection outside the sunspot 
\citep{tho02b,wei04}.
In our scenario the magnetic field aligned flows are
also siphon flows. The length of the loops is also small
so that descending footpoints are often in the penumbrae.
The difference of magnetic field 
between the hot and cold penumbral components 
drives the Evershed flow.
Our scenario includes (slow) motions across field
lines from the hot to the cold penumbral filaments.
The downflowing footpoints are continuously
pumped down by the drag of magnetic downdrafts within the penumbra, 
rather than outside the sunspot.

	To the best of our knowledge,
	interchange convection was put forward
as a mechanism of energy transport
by \citet{sch91} and \citet{jah94}.
The transfer is produce by uplifting 
hot fluxtubes from the magnetopause\footnote{
Boundary layer that separates the sunspot from the 
non-magnetic surrounding, and stays 
below the observable layers.}
to the photosphere, where they cool down releasing energy.
It has been modeled
in detail by \citeauthor{sch02d} and collaborators
\citep{sch98a,sch98b,sch02d}. 
The uplifting of hot magnetic fluxtubes is due to a convective
instability (hot and so lightweight fluxtubes tend to
rise).
In our case, however, the presence of 
vigorous downdrafts and the conservation of mass
force hot material to rise next to the 
downdraft, producing
bright filaments surrounding
the dark cores.
As far as the similarities with the interchange
convection is concerned,
we also have strong flows along field lines, 
with the mass cooling down as it radiates away.
The differences being the length of the arches
(shorter in our scenario), and the submergence of the
cold footpoint, which does return to the solar
sub-photosphere in the penumbra. (These differences
are not so obvious  for the  
sea-serpent fluxtube in \citealt{sch02d}.)

	Again some of the aspects of our
scenario are similar to the convective rolls
discussed by \citet{dan61}. In our case, however, we have 
strong motions
along field lines. Moreover, it is not clear whether
the submerging footpoint will return to 
the photosphere in a cyclic motions like 
the original \citeauthor{dan61}'s rolls.  
It may be closer to the paradigm in non-magnetic
convection \citep{stei98}, which is highly non-local
with strong  downdrafts connecting many different
pressure scale heights, 
and where only a small fraction of the rising 
hot material actually reaches the photosphere.
The degree of similarity would be established
only when a realistic numerical MHD modeling
of the process is available.

%

%

\section{Conclusions and empirical tests}\label{conclusions}

	We describe  a penumbral magnetic field
formed by short magnetic loops most of which
return to the sub-photosphere within
the sunspot boundary\footnote{Open field lines are also
needed to account for the large scale structure
of the sunspot magnetic fields.}
 (Fig.~\ref{cartoon2}). 
Matter flowing along
magnetic field lines would give rise to the
Evershed effect. This flow along field lines
would also be responsible for the
convective transport of heat in penumbrae.
Cold downflows of many loops are aligned in a sort
of lane of downdrafts whose observational
counterpart would be 
the dark cores in penumbral filaments found by
Scharmer et al. (2002).
To some extent, the scenario
resembles the convective rolls put forward by 
Danielson some 40 years ago, except that (a) 
the mass also flows along field lines, and (b)
the mass may not re-emerge
after submergence. It is also
akin to the interchange convection 
\citep[e.g.][]{sch02d}, where flows
along field lines transport heat from below
and give rise to the Evershed effect phenomena.
In our case, however,  the 
uprise of the hot tubes is induced
by the presence of downdrafts and the need to satisfy
mass conservation (as in the non-magnetic
granular convection, e.g., \citealt{ras03}). 
This downdrafts pump down the cool footpoints
of the loops as numerical models have shown to
occur with the downdrafts outside the 
penumbra \citep{wei04}.
Finally, our scenario also owes some properties to the
siphon flow model \citep{mey68,tho93}. A gas
pressure difference between the loop footpoints drives
the  flows, but this difference is set by the 
magnetic field strength difference between
the hot and the cold penumbral fibrils,
rather than from differences between 
the penumbra and magnetic concentrations
around the sunspot.

This scenario seems to be compatible with most observations
existing in the literature, in particular, the ubiquitous
downflows and the return of 
magnetic flux found by \paperiii\ (see \S~\ref{qua_obs}). 
However,
a difficulty remains. The upflows and downflows found
by \paperiii\ are not fast enough to transport
the flux of radiation emanating
from penumbrae (\S~\ref{not_enough}). 
A way out would
be a bias in the observations.
Insufficient
spatial resolution leads to the cancellation
of upflows and
downflows and  provides only 
a small residual (\S~\ref{observations}). Fortunately, the
presence of sizeable upflows and downflows in the
penumbra is a conjecture that can be tested.
Large upflows and
downflows must show up upon improving
of the angular resolution. Several specific
tests which will allow to confirm or falsify 
our scenario are:
\begin{itemize}
\item Diffraction limited high resolution spectroscopy with 
1-meter class  telescopes  must
show a correlation between 
brightness and velocity. The sign of the correlation
cannot change
from the center-side penumbra to the limb-side penumbra.
(This correlation is actually observed; see
\itm{correlation} in \S~\ref{observations}. Increasing the 
angular resolution
would increase the amplitude of the flows up to
a velocity of the order of 1~km~s$^{-1}$;
\S~\ref{not_enough}.)
\item The molecular spectra trace the cold 
penumbral component \citep[e.g.][]{pen03}. They should show
a global downflow of the order of
1~km~s$^{-1}$, no matter the angular resolution
of the observation. 
\item High angular resolution longitudinal magnetograms 
in sunspots close to the solar disk center
should show both positive and negative
polarities. In particular,
the dark cores 
are associated with magnetic field lines returning to
the sub-photosphere and, therefore, they should have a
polarity opposite to the umbra. 
\item Local correlation tracking techniques
\citep{nov88} applied to series of penumbral images
should reveal
bright features moving toward dark filaments. 
Such motion is masked
by the tendency of the bright plasma to become 
dark and transparent
as it approaches a dark core.
A global trend is to be expected, though.
Such trend seems to be present in high
resolution penumbral images \citep{mar05}.
\end{itemize}

Most of the penumbra
observed by \citet{sch02} does not show obvious 
dark cores (e.g., Fig.~\ref{narrow}), which implies that many
cores are probably not spatially resolved
with the present angular resolution.
The observed ones represent large
and lonely specimens 
among the family of penumbral downdrafts. Resolving the
smallest ones is not only a question of
improving the angular resolution. They 
are probably close to the optically-thin limit, where
the radiative transfer smearing 
modulates and 
reduces the contrast between hot upflows and
cold downflows.
We conjecture that the light bridges (LB)
are scaled up 
versions of the basic building blocks forming
the penumbral
structures (two paired convective rolls sharing a single
downdraft; \S~\ref{scenario}).  
The magnetic micro-structure may be larger in these LBs
since the presence
of aligned bright points and dark lanes is not so elusive
as the dark cores in penumbral filaments
(Fig.~\ref{light}). If the LB magnetic
structure is larger,  the observations
to falsify our penumbral scenario may be simpler to 
try with LBs and so, they represent an ideal target
for a first trial.


\acknowledgements
In a paper for non-specialists, \citet{pri03}
speculates 
{\em Are the bright filaments doubly
convective rolls with a dark core that are cooling and
sinking?} This question seems to advance
the scenario advocated here.
Our paper took form during a short visit paid
to the observatory of Meudon, France,  thanks to
the kind invitation of
C. Briand.
Images in Figs.~\ref{narrow}, \ref{coresfig}, and \ref{light}  have been
generously provided by the Institute for Solar Physics
of the Royal Swedish Academy of Sciences. They were
obtained with the Swedish Vacuum Solar Telescope
(SVST) and the Swedish Solar Telescope
(SST) operated in the Spanish Observatorio del
Roque de Los Muchachos (La Palma). 
Thanks are due J.~A.~Bonet and I.~M\'arquez for 
discussions on some of the ideas presented
in the paper, in particular, the local correlation
tracking test in \S~\ref{conclusions}. 
The work has partly been funded by the Spanish Ministry of Science
and Technology, 
project AYA2004-05792, as well as by
the EC contract HPRN-CT-2002-00313
%
%

\begin{thebibliography}{73}
\expandafter\ifx\csname natexlab\endcsname\relax\def\natexlab#1{#1}\fi

\bibitem[{{Arena} {et~al.}(1990){Arena}, {Landi degl'Innocenti}, \&
  {Noci}}]{are90}
{Arena}, P., {Landi degl'Innocenti}, E., \& {Noci}, G. 1990, \solphys, 129, 259

\bibitem[{{Beckers} \& {Schr{\" o}ter}(1969)}]{bec69c}
{Beckers}, J.~M., \& {Schr{\" o}ter}, E.~H. 1969, \solphys, 10, 384

\bibitem[{{Bellot Rubio}(2004)}]{bel04}
{Bellot Rubio}, L.~R. 2004, Rev. Mod. Astron., 17, 21

\bibitem[{{Bellot Rubio} {et~al.}(2003){Bellot Rubio}, {Balthasar}, {Collados},
  \& {Schlichenmaier}}]{bel03b}
{Bellot Rubio}, L.~R., {Balthasar}, H., {Collados}, M., \& {Schlichenmaier}, R.
  2003, \aap, 403, L47

\bibitem[{{Bonet} {et~al.}(2005){Bonet}, {M\'arquez}, {Muller}, {Sobotka}, \&
  {Roudier}}]{bon04}
{Bonet}, J.~A., {M\'arquez}, I., {Muller}, R., {Sobotka}, M., \& {Roudier}, T.
  2005, \aap, in press

\bibitem[{{Bovelet} \& {Wiehr}(2003)}]{bov03}
{Bovelet}, B., \& {Wiehr}, E. 2003, \aap, 412, 249

\bibitem[{{Bumba}(1960)}]{bum60}
{Bumba}, V. 1960, Izv.~Crim.~Astrophys.~Obs., 23, 253

\bibitem[{{Danielson}(1961)}]{dan61}
{Danielson}, R.~E. 1961, \apj, 134, 289

\bibitem[{{Garc\'\i a de la Rosa}(1987)}]{gar87}
{Garc\'\i a de la Rosa}, J.~I. 1987, in The Role of Fine-Scale Magnetic Fields
  on the Structure of the Solar Atmosphere, ed. E.-H. {Schr\"o ter},
  M.~{V\'aquez}, \& A.~A. {Wyller} (Cambridge: Cambridge University Press), 140

\bibitem[{{Golovko}(1974)}]{gol74}
{Golovko}, A.~A. 1974, \solphys, 37, 113

\bibitem[{{Grigorjev} \& {Katz}(1972)}]{gri72}
{Grigorjev}, V.~M., \& {Katz}, J.~M. 1972, \solphys, 22, 119

\bibitem[{{Hofmann} {et~al.}(1994){Hofmann}, {Deubner}, {Fleck}, \&
  {Schmidt}}]{hof94}
{Hofmann}, J., {Deubner}, F.-L., {Fleck}, B., \& {Schmidt}, W. 1994, \aap, 284,
  269

\bibitem[{{Hurlburt} {et~al.}(2000){Hurlburt}, {Matthews}, \&
  {Rucklidge}}]{hur00}
{Hurlburt}, N.~E., {Matthews}, P.~C., \& {Rucklidge}, A.~M. 2000, \solphys,
  192, 109

\bibitem[{{Illing} {et~al.}(1974{\natexlab{a}}){Illing}, {Landman}, \&
  {Mickey}}]{ill74a}
{Illing}, R. M.~E., {Landman}, D.~A., \& {Mickey}, D.~L. 1974{\natexlab{a}},
  \aap, 35, 327

\bibitem[{{Illing} {et~al.}(1974{\natexlab{b}}){Illing}, {Landman}, \&
  {Mickey}}]{ill74b}
---. 1974{\natexlab{b}}, \aap, 37, 97

\bibitem[{{Jahn} \& {Schmidt}(1994)}]{jah94}
{Jahn}, K., \& {Schmidt}, H.~U. 1994, \aap, 290, 295

\bibitem[{{Johannesson}(1993)}]{joh93}
{Johannesson}, A. 1993, \aap, 273, 633

\bibitem[{{K\`alm\`an}(1991)}]{kal91}
{K\`alm\`an}, B. 1991, \solphys, 135, 299

\bibitem[{{Leka}(1997)}]{lek97}
{Leka}, K.~D. 1997, \apj, 484, 900

\bibitem[{{Lites} {et~al.}(1993){Lites}, {Elmore}, {Seagraves}, \&
  {Skumanich}}]{lit93}
{Lites}, B.~W., {Elmore}, D.~F., {Seagraves}, P., \& {Skumanich}, A. 1993,
  \apj, 418, 928

\bibitem[{{Lites} {et~al.}(2004){Lites}, {Scharmer}, {Berger}, \&
  {Title}}]{lit04}
{Lites}, B.~W., {Scharmer}, G.~B., {Berger}, T.~E., \& {Title}, A.~M. 2004,
  \solphys, 221, 65

\bibitem[{{Lites} \& {Skumanich}(1990)}]{lit90}
{Lites}, B.~W., \& {Skumanich}, A. 1990, \apj, 348, 747

\bibitem[{{Makita}(1986)}]{mak86}
{Makita}, M. 1986, \solphys, 106, 269

\bibitem[{{M\'arquez} {et~al.}(2005){M\'arquez}, {S\'anchez Almeida}, \&
  {Bonet}}]{mar05}
{M\'arquez}, I., {S\'anchez Almeida}, J., \& {Bonet}, J.~A. 2005, \apj, in
  preparation

\bibitem[{{Meyer} \& {Schmidt}(1968)}]{mey68}
{Meyer}, F., \& {Schmidt}, H.~U. 1968, Mitteilungen der Astronomischen
  Gesellschaft Hamburg, 25, 194

\bibitem[{{Meyer} {et~al.}(1977){Meyer}, {Schmidt}, \& {Weiss}}]{mey77}
{Meyer}, F., {Schmidt}, H.~U., \& {Weiss}, N.~O. 1977, \mnras, 179, 741

\bibitem[{{Montesinos} \& {Thomas}(1997)}]{mon97}
{Montesinos}, B., \& {Thomas}, J.~H. 1997, \nat, 390, 485

\bibitem[{{Moreno-Insertis} \& {Spruit}(1989)}]{mor89}
{Moreno-Insertis}, F., \& {Spruit}, H.~C. 1989, \apj, 342, 1158

\bibitem[{{November} \& {Simon}(1988)}]{nov88}
{November}, L.~J., \& {Simon}, G.~W. 1988, \apj, 333, 427

\bibitem[{{Osherovich}(1982)}]{osh82}
{Osherovich}, V.~A. 1982, \solphys, 77, 63

\bibitem[{{Parker}(1975)}]{par75}
{Parker}, E.~N. 1975, \solphys, 40, 291

\bibitem[{{Penn} {et~al.}(2003){Penn}, {Cao}, {Walton}, {Chapman}, \&
  {Livingston}}]{pen03}
{Penn}, M.~J., {Cao}, W.~D., {Walton}, S.~R., {Chapman}, G.~A., \&
  {Livingston}, W. 2003, \apjl, 590, L119

\bibitem[{{Priest}(2003)}]{pri03}
{Priest}, E. 2003, Physics World, 16(2), 19

\bibitem[{{R\" uedi} {et~al.}(1999){R\" uedi}, {Solanki}, \& {Keller}}]{rue99}
{R\" uedi}, I., {Solanki}, S.~K., \& {Keller}, C.~U. 1999, \aap, 348, L37

\bibitem[{{Rast}(1995)}]{ras95}
{Rast}, M.~P. 1995, \apj, 443, 863

\bibitem[{{Rast}(2003)}]{ras03}
---. 2003, \apj, 597, 1200

\bibitem[{{Rimmele}(1995)}]{rim95}
{Rimmele}, T.~R. 1995, \apj, 445, 511

\bibitem[{{Rimmele}(1997)}]{rim97}
---. 1997, \apj, 490, 458

\bibitem[{{Rouppe van der Voort} {et~al.}(2004){Rouppe van der Voort}, {L{\"
  o}fdahl}, {Kiselman}, \& {Scharmer}}]{rou04}
{Rouppe van der Voort}, L.~H.~M., {L{\" o}fdahl}, M.~G., {Kiselman}, D., \&
  {Scharmer}, G.~B. 2004, \aap, 414, 717

\bibitem[{{S\'anchez Almeida}(1998)}]{san98a}
{S\'anchez Almeida}, J. 1998, \apj, 497, 967

\bibitem[{{S\'anchez Almeida}(2001)}]{san01c}
---. 2001, \apj, 556, 928

\bibitem[{{S\'anchez Almeida}(2005)}]{san04b}
---. 2005, \apj, 622, in press, (SA05, astro-ph/0412217)

\bibitem[{{S\'anchez Almeida} {et~al.}(1996){S\'anchez Almeida}, {Landi
  Degl'Innocenti}, {Mart\'\i nez Pillet}, \& {Lites}}]{san96}
{S\'anchez Almeida}, J., {Landi Degl'Innocenti}, E., {Mart\'\i nez Pillet}, V.,
  \& {Lites}, B.~W. 1996, \apj, 466, 537

\bibitem[{{S\'anchez Almeida} \& {Lites}(1992)}]{san92b}
{S\'anchez Almeida}, J., \& {Lites}, B.~W. 1992, \apj, 398, 359

\bibitem[{{S\'anchez Almeida} {et~al.}(1993){S\'anchez Almeida}, {Mart\'\i nez
  Pillet}, {Trujillo Bueno}, \& {Lites}}]{san93b}
{S\'anchez Almeida}, J., {Mart\'\i nez Pillet}, V., {Trujillo Bueno}, J., \&
  {Lites}, B.~W. 1993, in ASP Conf. Ser., Vol.~46, The Magnetic and Velocity
  Fields of Solar Active Regions, ed. H.~{Zirin}, G.~{Ai}, \& H.~{Wang} (San
  Francisco: ASP), 192

\bibitem[{{Scharmer} {et~al.}(1985){Scharmer}, {Brown}, {Pettersson}, \&
  {Rehn}}]{sch85}
{Scharmer}, G.~B., {Brown}, D.~S., {Pettersson}, L., \& {Rehn}, J. 1985, \ao,
  24, 2558

\bibitem[{{Scharmer} {et~al.}(2002){Scharmer}, {Gudiksen}, {Kiselman}, {L{\"
  o}fdahl}, \& {van der Voort}}]{sch02}
{Scharmer}, G.~B., {Gudiksen}, B.~V., {Kiselman}, D., {L{\" o}fdahl}, M.~G., \&
  {van der Voort}, L.~H.~M.~R. 2002, \nat, 420, 151

\bibitem[{{Schlichenmaier}(2002)}]{sch02d}
{Schlichenmaier}, R. 2002, Astron. Nachr., 323, 303

\bibitem[{{Schlichenmaier} {et~al.}(1999){Schlichenmaier}, {Bruls}, \& {Sch{\"
  u}ssler}}]{sch99}
{Schlichenmaier}, R., {Bruls}, J.~H.~M.~J., \& {Sch{\" u}ssler}, M. 1999, \aap,
  349, 961

\bibitem[{{Schlichenmaier} \& {Collados}(2002)}]{sch02b}
{Schlichenmaier}, R., \& {Collados}, M. 2002, \aap, 381, 668

\bibitem[{{Schlichenmaier} {et~al.}(1998{\natexlab{a}}){Schlichenmaier},
  {Jahn}, \& {Schmidt}}]{sch98a}
{Schlichenmaier}, R., {Jahn}, K., \& {Schmidt}, H.~U. 1998{\natexlab{a}},
  \apjl, 493, L121

\bibitem[{{Schlichenmaier} {et~al.}(1998{\natexlab{b}}){Schlichenmaier},
  {Jahn}, \& {Schmidt}}]{sch98b}
---. 1998{\natexlab{b}}, \aap, 337, 897

\bibitem[{{Schlichenmaier} \& Schmidt(2000)}]{sch00}
{Schlichenmaier}, R., \& Schmidt, W. 2000, \aap, 358, 1122

\bibitem[{{Schlichenmaier} \& {Solanki}(2003)}]{sch03f}
{Schlichenmaier}, R., \& {Solanki}, S.~K. 2003, \aap, 411, 257

\bibitem[{Schmidt(1991)}]{sch91}
Schmidt, H.~U. 1991, Geophys. Astrophys. Fluid Dyn., 62, 249

\bibitem[{{Schmidt} {et~al.}(1986){Schmidt}, {Spruit}, \& {Weiss}}]{sch86b}
{Schmidt}, H.~U., {Spruit}, H.~C., \& {Weiss}, N.~O. 1986, \aap, 158, 351

\bibitem[{{Schmidt} \& {Fritz}(2004)}]{sch04}
{Schmidt}, W., \& {Fritz}, G. 2004, \aap, 421, 735

\bibitem[{{Schmidt} \& {Schlichenmaier}(2000)}]{sch00b}
{Schmidt}, W., \& {Schlichenmaier}, R. 2000, \aap, 364, 829

\bibitem[{{Sheeley}(1972)}]{she72}
{Sheeley}, N.~R. 1972, \solphys, 25, 98

\bibitem[{{Solanki}(2003)}]{sol03}
{Solanki}, S.~K. 2003, \aapr, 11, 153

\bibitem[{{Spruit}(1987)}]{spr87}
{Spruit}, H.~C. 1987, in The Role of Fine-Scale Magnetic Fields on the
  Structure of the Solar Atmosphere, ed. E.-H. {Schr\"o ter}, M.~{V\'aquez}, \&
  A.~A. {Wyller} (Cambridge: Cambridge University Press), 199

\bibitem[{{Stanchfield} {et~al.}(1997){Stanchfield}, {Thomas}, \&
  {Lites}}]{sta97}
{Stanchfield}, D.~C.~H., {Thomas}, J.~H., \& {Lites}, B.~W. 1997, \apj, 477,
  485

\bibitem[{{Stein} \& {Nordlund}(1998)}]{stei98}
{Stein}, R.~F., \& {Nordlund}, {\AA}. 1998, \apj, 499, 914

\bibitem[{{Stix}(1991)}]{sti91}
{Stix}, M. 1991, The Sun (Berlin: Springer-Verlag)

\bibitem[{{Thomas} \& {Montesinos}(1993)}]{tho93}
{Thomas}, J.~H., \& {Montesinos}, B. 1993, \apj, 407, 398

\bibitem[{{Thomas} \& {Weiss}(1992)}]{tho92}
{Thomas}, J.~H., \& {Weiss}, N.~O. 1992, in NATO ASI Ser., Vol. 375, Sunspots.
  Theory and Observations, ed. J.~H. {Thomas} \& N.~O. {Weiss} (Dordrecht:
  Kluwer), 3

\bibitem[{{Thomas} \& {Weiss}(2004)}]{tho04}
{Thomas}, J.~H., \& {Weiss}, N.~O. 2004, \araa, 42, 517

\bibitem[{{Thomas} {et~al.}(2002){Thomas}, {Weiss}, {Tobias}, \&
  {Brummell}}]{tho02b}
{Thomas}, J.~H., {Weiss}, N.~O., {Tobias}, S.~M., \& {Brummell}, N.~H. 2002,
  \nat, 420, 390

\bibitem[{{Title} {et~al.}(1993){Title}, {Frank}, {Shine}, {Tarbell}, {Topka},
  {Scharmer}, \& {Schmidt}}]{tit93}
{Title}, A.~M., {Frank}, Z.~A., {Shine}, R.~A., {Tarbell}, T.~D., {Topka},
  K.~P., {Scharmer}, G., \& {Schmidt}, W. 1993, \apj, 403, 780

\bibitem[{{Tritschler} {et~al.}(2004){Tritschler}, {Schlichenmaier}, {Bellot
  Rubio}, \& {the KAOS Team}}]{tri04}
{Tritschler}, A., {Schlichenmaier}, R., {Bellot Rubio}, L.~R., \& {the KAOS
  Team}. 2004, \aap, 415, 717

\bibitem[{{Weiss} {et~al.}(2004){Weiss}, {Thomas}, {Brummell}, \&
  {Tobias}}]{wei04}
{Weiss}, N.~O., {Thomas}, J.~H., {Brummell}, N.~H., \& {Tobias}, S.~M. 2004,
  \apj, 600, 1073

\bibitem[{{Westendorp Plaza} {et~al.}(1997){Westendorp Plaza}, {del Toro
  Iniesta}, {Ruiz Cobo}, {Mart\'\i nez Pillet}, {Lites}, \&
  {Skumanich}}]{wes97}
{Westendorp Plaza}, C., {del Toro Iniesta}, J.~C., {Ruiz Cobo}, B., {Mart\'\i
  nez Pillet}, V., {Lites}, B.~W., \& {Skumanich}, A. 1997, \nat, 389, 47

\bibitem[{{Wiehr}(1995)}]{wie95}
{Wiehr}, E. 1995, \aap, 298, L17

\end{thebibliography}

%
%
\appendix
\section{Reduction of velocity signals due to
insufficient spatial resolution}\label{appa}

Assume the
true vertical velocities to  have a sinusoidal
pattern with a period $l$, 
\begin{equation}
U_z(x,y)=U_z(x)=U_{z0}\sin(2\pi x/l).
\end{equation}
The symbols $x$ and $y$ stand for the two spatial coordinates,
with $y$ along the filaments where
$U_z$ is approximately constant. 
The observational
procedure limits the resolution 
as given by a point spread function (PSF) $P(x,y)$,
\begin{equation}
<U_z(x,y)>=<U_z(x)>=\int_{-\infty}^\infty\int_{-\infty}^\infty
	P(x'-x,y'-y)U_z(x')
	dx'dy',
\end{equation}
with $<U_z(x)>$
the observed vertical velocity.
We assume $P$ to be a Gaussian function,
\begin{equation}
P(x,y)={{4\ln 2}\over{\pi L^2}}\exp-\Big[{{4\ln 2}\over{L^2}}(x^2+y^2)\Big],
\end{equation}
with  full width half maximum $L$.
Then the damping of the true signals due to this 
PSF is
\begin{equation}
<U_z(x,y)>/U_z(x,y)=\exp-\Big[{{\pi^2}\over{4\ln 2}}{{L^2}\over{l^2}}\Big], 
\end{equation}
or
\begin{equation}
<U_z(x)>/U_z(x)\simeq\cases{
	4.1\times 10^{-1}&if $L = l/2$,\cr
	2.8\times 10^{-2}&if $L=l$,\cr
	3.3\times 10^{-4}&if $L = 3 l/2.$
		} 
\end{equation}
The case $l=L$ corresponds to an observation with
0\farcs 25 spatial resolution
if the velocities follow the pattern
of penumbral filaments with dark
cores found by \citeauthor{sch02}~(\citeyear{sch02}; $l\simeq 180$~km).

%
%
\end{document}